\documentstyle[11pt,epsf,subequation]{article}
\newcommand{\postscript}[2]
   {\setlength{\epsfxsize}{#2\hsize}
   \centerline{\epsfbox{#1}}}
\begin{document}
\def\theequation {\thesection.\arabic{equation}}
\makeatletter\@addtoreset {equation}{section}\makeatother
\title{\bf Multiple permanent-wave trains in nonlinear systems}
\author{Jianke Yang \\
 Department of Mathematics and Statistics \\
The University of Vermont \\
16 Colchester Avenue \\
Burlington, VT 05401 \\
Tel.: 802-656-4314 \\
Fax:  802-656-2552 \\
E-mail: jyang@emba.uvm.edu }
\date{    }
\maketitle
\thispagestyle{empty}
{\bf Abstract}

Multiple permanent-wave trains in nonlinear systems are constructed 
by the asymptotic tail-matching method. Under some general assumptions, 
simple criteria for the construction are presented. Applications to fourth-order
systems and coupled nonlinear Schr\"odinger equations are discussed. 

\section{Introduction}

Nonlinear wave systems have been studied for a few decades. Much progress has been 
made on integrable equations where the inverse scattering transform method can
be applied \cite{ablowitz}. For non-integrable equations, the general
analytical treatment has been elusive so far and will likely remain so in the near 
future. A less ambitious goal, then, would be to generally study the permanent 
waves in non-integrable systems. Such waves often contain valuable information on
the system's general solution behaviors. An interesting fact is that, in 
many non-integrable systems, simple permanent waves can be matched together and
form multiple permanent-wave trains (\cite{yang}, \cite{balmforth}, \cite{buffoni},
\cite{vandervorst}, \cite{klauder}, \cite{akylas}, etc.). 
If solitary waves exist, multiple solitary-wave trains can be constructed by 
a perturbation method proposed by Karpman and Solov'ev \cite{karpman}
and Gorshkov and Ostrovsky \cite{gorshkov} (see \cite{balmforth}). 
If permanent waves with exponentially
decaying and oscillating tails are present,  
the existence of countably infinite multiple
permanent-wave trains has been proved for certain types of nonlinear systems
by variational methods (\cite{buffoni}, \cite{vandervorst}). 
In this paper, if a nonlinear wave system allows permanent waves 
which exponentially approach a constant at 
infinity, we will construct widely-separated multiple permanent-wave trains by a 
new and general method, namely, the asymptotic tail-matching method. 
This method is stimulated in part by another matching method for non-local
solitary waves (see \cite{klauder} and \cite{akylas}). 
Under some general assumptions, we will show that an arbitrary number of permanent
waves can be matched together and form multiple permanent-wave trains if and only
if the exponential tails of these permanent waves satisfy certain simple algebraic
conditions. These conditions will also determine the spacings between adjacent 
permanent waves if such matching takes place. 
This asymptotic tail-matching method differs from 
Karpman et al's perturbation method in two major aspects. First, it 
can be applied directly to the matching of kink and anti-kink type permanent waves. 
Second, its results are explicit, simple and insightful.
As applications of these general results, we will discuss fourth-order
systems and the coupled nonlinear Schr\"odinger equations. For fourth-order 
systems which allow permanent waves exponentially and oscillatorily approaching
a constant at infinity, we will show that countably infinite multiple
permanent-wave trains exist and can be readily constructed. Thus the results in
\cite{buffoni} and \cite{vandervorst} are reproduced. For 
the coupled nonlinear Schr\"odinger equations, we will show that 
countably infinite multiple solitary-wave trains can be constructed 
in a large portion of the parameter space. Numerical results will also be presented
and compared with the theoretical predictions when appropriate. 

\section{Construction of multiple permanent-wave trains}

We consider a general nonlinear wave system
\begin{equation} \label{pde}
F(U, D_{x}, D_{t})=0,
\end{equation}
where $U$ is the unknown vector variable, and $F$ is a nonlinear vector function. 
Suppose it allows permanent waves of certain form which, when substituted into Eq. 
(\ref{pde}), reduces it into an autonomous complex system of first-order nonlinear
ordinary differential equations
\begin{equation}\label{ode}
d\Phi/dx=G(\Phi),
\end{equation}
where $\Phi(x)$ is a $n$-component vector variable. If Eq. (\ref{ode})
has permanent wave solutions which exponentially approach a constant at infinity, 
then we next will develop a new method to determine if those permanent waves can
be matched together and form widely-separated 
multiple permanent-wave trains or not. The idea is 
to perturb each permanent wave such that the exponential tails of each perturbed wave
match those of the adjacent permanent waves. We first discuss the matching of solitary
waves, followed by that of general permanent waves. 

\subsection{Solitary-wave trains}
Suppose Eq. (\ref{ode}) allows solitary waves $\Phi(x)$ which exponentially decay
to zero as $|x|\rightarrow \infty$. We make the following general assumptions: 
\begin{enumerate}
\item[A1.] the eigenvalues of the constant (Jacobian) 
matrix $\nabla G(0)$ all have non-zero real parts;
\item[A2.] for any solitary wave $\Phi(x)$, the linear behavior dominates at
infinity, i.e., as $x \rightarrow \infty$ or $-\infty$, 
$\Phi(x)$ approaches a solution of the linear equation
\begin{equation} \label{equinf}
d\tilde{\Phi}/dx=\nabla G(0)\tilde{\Phi};
\end{equation}
\item[A3.] for any solitary wave $\Phi(x)$, 
the linearized equation of (\ref{ode}) around $\Phi(x)$ 
\begin{equation} \label{linear}
d\tilde{\Phi}/dx=\nabla G(\Phi)\tilde{\Phi}
\end{equation}
and its adjoint equation
\begin{equation}  \label{adjoint1}
-d\Psi/dx=\nabla G^{*\mbox{\tiny\it T}}(\Phi)\Psi
\end{equation}
each have a single linearly independent localized solution. Here ``T'' represents
the transpose and ``*'' the complex conjugate. 
\end{enumerate}

Remark: Since Eq. (\ref{ode}) is autonomous, any spatial translation of $\Phi(x)$
is still (\ref{ode})'s solution. Therefore Eq. (\ref{linear}) always has a nontrivial
localized solution $d\Phi(x)/dx$. The requirement for Eq. (\ref{linear}) is just that
$d\Phi(x)/dx$ is its single linearly independent localized solution. 
This can be guaranteed if $\Phi(x)$  
is isolated in $H^{1}(R, R^{2})$ up to spatial 
translations (the so-called non-degenerency
condition in some of the literature. See \cite{buffoni}). 

We also introduce the following notations. In view of assumption A1, 
let us denote $\nabla G(0)$'s eigenvalues 
as $\lambda_{1}, \lambda_{2}, \dots, \lambda_{n}$ where
\begin{equation}
\mbox{Re}(\lambda_{1})\le \mbox{Re}(\lambda_{2})\le \dots \le \mbox{Re}(\lambda_{s})<0<
\mbox{Re}(\lambda_{s+1})\le \mbox{Re}(\lambda_{s+2})\le 
\dots \le \mbox{Re}(\lambda_{n}).
\end{equation}
Suppose for an eigenvalue $\lambda$, $\nabla G(0)$ has a chain of eigenvector and 
generalized eigenvectors $v_{i}\; (i=1,\dots, l)$ such that 
\begin{equation}
(\nabla G(0)-\lambda I)v_{1}=0, 
\end{equation}
\begin{equation}
(\nabla G(0)-\lambda I)v_{i+1}=v_{i}, \hspace{0.6cm} i=1, \dots, l-1. 
\end{equation}
Define the polynomial functions $\xi_{i}(x)\; (i=1,\dots, l)$ as
\begin{equation} \label{ksix}
\xi_{i}(x)=v_{i}+xv_{i-1}+\dots+\frac{x^{i-1}}{(i-1)!}v_{1}, \hspace{1cm} i=1,\dots, l,
\end{equation}
then 
\begin{equation}
\xi'_{i+1}(x)=\xi_{i}(x),
\end{equation}
and $\{\xi_{i}(x)e^{\lambda x}, \; i=1,\dots, l\}$ form a chain of linearly independent
solutions of Eq. (\ref{equinf}). According to the theory of linear differential 
equations with constant coefficients, we can find such chains of solutions which
together form a fundamental set of solutions of Eq. (\ref{equinf}). Thus according to 
assumption A2, we have 
\begin{equation}
\Phi(x) \longrightarrow 
\left\{ \begin{array}{l} 
\sum_{i=1}^{s}c_{i}\xi_{i}(x)e^{\lambda_{i}x}, \hspace{0.8cm} x\rightarrow \infty, \\
\sum_{i=s+1}^{n}c_{i}\xi_{i}(x)e^{\lambda_{i}x}, \hspace{0.4cm} x\rightarrow -\infty,
\end{array} \right.
\end{equation}
where $c_{i}\; (i=1, \dots, n)$ are complex constants. We point out that in the special
case where $\nabla G(0)$ has $n$ linearly independent eigenvectors, $\{\xi_{i},\: 
i=1, \dots, n\}$ are just
those constant eigenvectors.
The fundamental matrix of the adjoint equation (\ref{adjoint1}) at infinity is 
\begin{equation}
[\eta_{1}e^{-\lambda_{1}^{*}x} \;\; \eta_{2}e^{-\lambda_{2}^{*}x}\;\; \dots \;\;
\eta_{n}e^{-\lambda_{n}^{*}x}],
\end{equation}
where 
\begin{equation}
[\eta_{1}\;\: \eta_{2}\;\: \dots \;\: \eta_{n}]=
\{[\xi_{1}\;\: \xi_{2}\;\:\dots \;\:
\xi_{n}]^{-1}\}^{*\mbox{\tiny\it T}}. 
\end{equation}
Note that for $1\le i, j\le n$, 
\begin{equation}
\xi_{i}(x)\cdot \eta^{*}_{j}(x)=\left\{
\begin{array}{l}
1, \hspace{0.5cm} i=j,  \\
0, \hspace{0.5cm} i\ne j. 
\end{array} \right.
\end{equation} 
Thus for the single linearly independent 
localized solution $\Psi(x)$ of Eq. (\ref{adjoint1}), we have
\begin{equation}
\Psi(x) \longrightarrow 
\left\{ \begin{array}{l} 
\sum^{s}_{i=1}d_{i}\eta_{i}(x)e^{-\lambda^{*}_{i}x}, 
\hspace{1.5cm} x\rightarrow -\infty, \\
\sum^{n}_{i=s+1}d_{i}\eta_{i}(x)e^{-\lambda^{*}_{i}x}, 
\hspace{1.1cm} x\rightarrow \infty,
\end{array} \right.
\end{equation}
where $d_{i}\; (i=1, \dots, n)$ are complex constants. 

Now suppose $\{\Phi^{(1)}, \Phi^{(2)}, \dots, \Phi^{(N)}\}$ are $N$ solitary 
waves of Eq. (\ref{ode}) with
\begin{equation} \label{asym}
\Phi^{(k)}(x) \longrightarrow 
\left\{ \begin{array}{l} 
\sum_{i=1}^{s}c_{i}^{(k)}
\xi_{i}(x)e^{\lambda_{i}x}, \hspace{0.7cm} x\rightarrow \infty, \\
\sum_{i=s+1}^{n}c_{i}^{(k)}
\xi_{i}(x)e^{\lambda_{i}x}, \hspace{0.3cm} x\rightarrow -\infty.
\end{array} \right.
\end{equation}
For each $\Phi^{(k)}$, the single linearly independent localized solution $\Psi^{(k)}$ 
of the adjoint equation
\begin{equation} \label{adjoint2}
-d\Psi^{(k)}/dx=\nabla G^{*\mbox{\tiny\it T}}(\Phi^{(k)})\Psi^{(k)}
\end{equation}
has the following asymptotic behavior at infinity: 
\begin{equation}
\Psi^{(k)}(x) \longrightarrow 
\left\{ \begin{array}{l} 
\sum^{s}_{i=1}d^{(k)}_{i}\eta_{i}(x)e^{-\lambda^{*}_{i}x}, \hspace{1.4cm} 
x\rightarrow -\infty, \\
\sum^{n}_{i=s+1}d^{(k)}_{i}\eta_{i}(x)e^{-\lambda^{*}_{i}x}, 
\hspace{1cm} x\rightarrow \infty. 
\end{array} \right.
\end{equation}
Consider a new solitary wave which looks like a superposition of the above $N$ solitary
waves $\{\Phi^{(k)}\}$ 
widely separated, with the $k$-th wave $\Phi^{(k)}$ located at $x=x_{k}$
$(k=1, 2, \dots, N)$. Let
\begin{equation}
x_{1} < x_{2} < \dots < x_{N},
\end{equation}
and denote 
\begin{equation} \label{deltak}
\triangle_{k}=x_{k+1}-x_{k}\; (\gg 1), \hspace{0.3cm} k=1, 2, \dots, N-1.
\end{equation}
We will call this new solitary wave
as a $N$-pulse wavetrain. It can be constructed explicitly by the following theorem.
\newtheorem{guess}{Theorem}
\begin{guess}
Under the assumptions A1, A2, A3 and the above notations, the $N$ solitary waves
$\{\Phi^{(1)}, \dots, \Phi^{(N)}\}$ can match each other and form a widely-separated
$N$-pulse wavetrain if and only if the spacings $\triangle_{k} \;(\gg 1)\; 
(k=1, \dots, N-1)$
asymptotically satisfy the following $N$ conditions
\begin{subequations} \label{formula1}
\begin{equation}   
\sum^{n}_{j=s+1}c_{j}^{(2)}d_{j}^{(1)*}e^{-\lambda_{j}\triangle_{1}}=0,
\end{equation}
\begin{equation}
\sum^{s}_{j=1}c_{j}^{(k-1)}d_{j}^{(k)*}e^{\lambda_{j}\triangle_{k-1}}
=\sum^{n}_{j=s+1}c_{j}^{(k+1)}d_{j}^{(k)*}e^{-\lambda_{j}\triangle_{k}},
\hspace{0.5cm} (2\le k\le N-1),
\end{equation}
\begin{equation}
\sum^{s}_{j=1}c_{j}^{(N-1)}d_{j}^{(N)*}e^{\lambda_{j}\triangle_{N-1}}=0.
\end{equation}
\end{subequations}
The relative errors in Eqs. (\ref{formula1}) are exponentially small with the spacings. 
\end{guess}
We will prove this theorem by the asymptotic tail-matching method to be developed next. 

Proof: Suppose such a $N$-pulse wavetrain $\Phi(x)$ exists. Then 
around the $k$-th wave $(2\le k\le N-1)$, the solution is 
\begin{equation} \label{phi}
\Phi(x)=\Phi^{(k)}(x-x_{k})+\tilde{\Phi}^{(k)}(x-x_{k}),
\end{equation}
where $\tilde{\Phi}^{(k)}\ll 1$. The linearized equation for $\tilde{\Phi}^{(k)}$ is
\begin{equation} \label{linear1}
d\tilde{\Phi}^{(k)}(x)/dx=\nabla G(\Phi^{(k)}(x)) \; \tilde{\Phi}^{(k)}(x).
\end{equation}
According to assumption A3, Eq. (\ref{linear1}) has a single linearly independent localized solution
which is $d\Phi^{(k)}(x)/dx$. 
From Eq. (\ref{asym}) we get 
\begin{equation} \label{dphi}
d\Phi^{(k)}(x)/dx \longrightarrow \sum^{s}_{i=1}c_{i}^{(k)}(\xi'_{i}(x)
+\lambda_{i}\xi_{i}(x))\;e^{\lambda_{i}x}, \hspace{0.5cm}
x\rightarrow \infty.
\end{equation}
Clearly not all the $c_{i}^{(k)}$'s $(i=1,\dots, s)$ are equal to zero. Without
loss of generality, we assume that $c_{1}^{(k)}\ne 0$. Then we denote the other $n-1$ 
solutions of Eq. (\ref{linear1}) as $\tilde{\Phi}_{j}^{(k)}\; (j=2,\dots, n)$. We
require that 
\begin{equation}
\tilde{\Phi}_{j}^{(k)}(x)\longrightarrow \xi_{j}(x)e^{\lambda_{j}x}, \hspace{0.5cm}
x\rightarrow \infty.
\end{equation}
As $x\rightarrow -\infty$, we generally have
\begin{equation}
\tilde{\Phi}_{j}^{(k)}(x)\longrightarrow \sum^{n}_{i=1}a_{ji}^{(k)}\xi_{i}(x)
e^{\lambda_{i}x}, \hspace{0.4cm} j=2, \dots, n,
\end{equation}
where $a_{ji}^{(k)}$ are constants. 
Since $c_{1}^{(k)}\ne 0$, these $n$ solutions $\{d\Phi^{(k)}/dx, \tilde{\Phi}_{2}^{(k)},
\dots, \tilde{\Phi}_{n}^{(k)}\}$ are linearly independent at $x$ equal to infinity, and
they form a fundamental set of solutions of Eq. (\ref{linear1}). 
Thus the general solution for $\tilde{\Phi}^{(k)}(x-x_{k})$ is
\begin{equation} \label{solution}
\tilde{\Phi}^{(k)}(x-x_{k})=h_{1} \frac{d\Phi^{(k)}}{dx}(x-x_{k})+\sum^{n}_{j=2}h_{j}
\tilde{\Phi}_{j}^{(k)}(x-x_{k}),
\end{equation}
where $h_{j}\; (j=1, \dots, n)$ are constants. 
The first term in (\ref{solution}) can be absorbed into $\Phi^{(k)}(x-x_{k})$ and 
cause a position shift to it. By normalization we make $h_{1}=0$.
When $x_{k}\ll x\ll x_{k+1}$, dropping the exponentially small terms, we get
\begin{equation} \label{phiasym1}
\tilde{\Phi}^{(k)}(x-x_{k}) \longrightarrow
\sum^{n}_{j=s+1}h_{j}\xi_{j}(x-x_{k})e^{\lambda_{j}
(x-x_{k})}.
\end{equation}
Similarly, when $x_{k-1}\ll x\ll x_{k}$, we have
\begin{equation} \label{phiasym2}
\tilde{\Phi}^{(k)}(x-x_{k}) \longrightarrow
\sum_{j=1}^{s}(\sum_{i=2}^{n}a_{ij}^{(k)}h_{i})\xi_{j}(x-x_{k})e^{\lambda_{j}(x-x_{k})}.
\end{equation}
The key idea in the asymptotic tail-matching method is that, in order for 
the matching to occur, we need to require that in the region 
$x_{k}\ll x\ll x_{k+1}$, $\tilde{\Phi}^{(k)}(x-x_{k})$'s exponentially growing
terms match the left tail of the right-hand wave $\Phi^{(k+1)}(x-x_{k+1})$;
in the region $x_{k-1}\ll x\ll x_{k}$, $\tilde{\Phi}^{(k)}(x-x_{k})$'s 
exponentially decaying terms match the right tail of the left-hand wave
$\Phi^{(k-1)}(x-x_{k-1})$. In the region $x_{k}\ll x\ll x_{k+1}$,
this requirement is 
\begin{equation} \label{cond1}
\sum_{j=s+1}^{n}h_{j}\xi_{j}(x-x_{k})e^{\lambda_{j}(x-x_{k})}
=\sum_{j=s+1}^{n}c_{j}^{(k+1)}\xi_{j}(x-x_{k+1})e^{\lambda_{j}(x-x_{k+1})};
\end{equation}
and in the region $x_{k-1}\ll x\ll x_{k}$, it is 
\begin{equation} \label{cond2}
\sum_{j=1}^{s}(\sum_{i=2}^{n}a_{ij}^{(k)}h_{i})
\xi_{j}(x-x_{k})e^{\lambda_{j}(x-x_{k})}
=\sum_{j=1}^{s}c_{j}^{(k-1)}\xi_{j}(x-x_{k-1})e^{\lambda_{j}(x-x_{k-1})}.
\end{equation}
We now need to select the constants $h_{j}\; (j=2,\dots, n)$ and spacings
$\triangle_{k}\; (k=1, \dots, N-1)$ so that the above two conditions are satisfied. 

First consider condition (\ref{cond1}). Recall that functions $\{\xi_{j}(x)\}$ are of 
the form (\ref{ksix}). If for $\lambda=\lambda_{m}\; (s+1\le m\le n)$, the chain
of such functions is $\{\xi_{m}, \dots, \xi_{m+l-1}\}$, where
$\xi_{m}$ is a constant vector and
\begin{equation}
\xi'_{j+1}(x)=\xi_{j}(x), \hspace{0.5cm} j=m, \dots, m+l-2. 
\end{equation}
Then we select $\{h_{m}, \dots, h_{m+j-1}\}$ from the equation
\begin{equation} \label{cond3}
\sum_{j=m}^{m+l-1}h_{j}\xi_{j}(x-x_{k})e^{\lambda(x-x_{k})}
=\sum_{j=m}^{m+l-1}c_{j}^{(k+1)}\xi_{j}(x-x_{k+1})e^{\lambda(x-x_{k+1})}.
\end{equation}
The right hand side of this equation is
\[ \sum_{j=m}^{m+l-1}c_{j}^{(k+1)}\xi_{j}(x-x_{k}-\triangle_{k})e^{\lambda(x-x_{k})}
e^{-\lambda \triangle_{k}}\]
\[=\sum_{j=m}^{m+l-1}c_{j}^{(k+1)}
(\sum_{i=0}^{j-m}\frac{(-\triangle_{k})^{i}}{i!}\frac{d^{i}\xi_{j}}{dx^{i}}(x-x_{k}))
e^{\lambda(x-x_{k})} e^{-\lambda \triangle_{k}}\]
\[=\sum_{j=m}^{m+l-1}\sum_{i=m}^{j}c_{j}^{(k+1)}\frac{(-\triangle_{k})^{j-i}}{(j-i)!}
\xi_{i}(x-x_{k})e^{\lambda(x-x_{k})} e^{-\lambda \triangle_{k}}\]
\begin{equation} 
=\sum_{i=m}^{m+l-1}(\sum_{j=i}^{m+l-1}\frac{(-\triangle_{k})^{j-i}}{(j-i)!}
c_{j}^{(k+1)})e^{-\lambda \triangle_{k}}\xi_{i}(x-x_{k})e^{\lambda(x-x_{k})}.
\end{equation}
Now we choose $h_{i}\; (i=m, \dots, m+l-1)$ to be
\begin{equation} \label{h1}
h_{i}=(\sum_{j=i}^{m+l-1}\frac{(-\triangle_{k})^{j-i}}{(j-i)!}
c_{j}^{(k+1)})e^{-\lambda \triangle_{k}}, \hspace{0.5cm} i=m, \dots, m+l-1, 
\end{equation}
then Eq. (\ref{cond3}) is valid. Repeating this procedure for the other chains of
\{$\xi_{j}(x)$\} functions in the form (\ref{ksix}), we can successfully select 
$h_{i}\; (i=s+1, \dots, n)$ so that condition (\ref{cond1}) is satisfied. 

Next consider condition (\ref{cond2}). Similar analysis shows that we can
reduce its right hand side to 
\begin{equation}
\sum^{s}_{j=1}c_{j}^{(k-1)}\xi_{j}(x-x_{k-1})e^{\lambda_{j}(x-x_{k-1})}
=\sum^{s}_{j=1}\alpha_{j}e^{\lambda_{j}\triangle_{k-1}}
\xi_{j}(x-x_{k})e^{\lambda_{j}(x-x_{k})},
\end{equation}
where $\alpha_{j}\; (j=1, \dots, s)$ are constants and determined by 
$c_{j}^{(k-1)}\; (j=1,\dots, s)$ and $\triangle_{k-1}$. 
Then condition (\ref{cond2}) becomes 
\begin{equation} \label{hh}
\sum^{s}_{i=2}a^{(k)}_{ij}h_{i}=\alpha_{j}e^{\lambda_{j}\triangle_{k-1}}-
\sum^{n}_{i=s+1}a^{(k)}_{ij}h_{i}, 
 \hspace{0.5cm} j=1, \dots, s.
\end{equation}
This is a linear system of $s$ equations for $s-1$ unknowns $h_{i}\; (i=2, \dots, s)$. 
We now show that the matrix $(a_{ij}^{(k)})_{s\times (s-1)}$ on the left side of
Eq. (\ref{hh}) has rank $s-1$. 
Consider the solution of Eq. (\ref{linear1}) 
\begin{equation}
T(x)=\sum_{j=2}^{s}p_{j}\tilde{\Phi}_{j}^{(k)}(x),
\end{equation}
where $p_{j}\; (j=2,\dots, s)$ are constants. Dropping exponentially small terms
we get
\begin{equation}
T(x) \longrightarrow 
\left\{ \begin{array}{l} 
0, \hspace{1cm} x\rightarrow \infty, \\
\sum_{j=1}^{s}(\sum^{s}_{i=2}a_{ij}^{(k)}p_{i})\xi_{j}(x)e^{\lambda_{j}x},
\hspace{0.3cm} x\rightarrow -\infty.
\end{array} \right.
\end{equation}
According to assumption A3, 
the only localized solution of Eq. (\ref{linear1}) is $d\Phi^{(k)}(x)/dx$. Moreover, 
$c_{1}^{(k)}$ in (\ref{dphi}) is non-zero. Thus $T(x)$ can not be a localized solution. 
In other words, the linear system of equations
\begin{equation}
\sum^{s}_{i=2}a_{ij}^{(k)}p_{i}=0, \hspace{1cm} j=1, \dots, s,
\end{equation}
has no non-trivial solutions for $p_{i}\; (i=1,\dots, s)$. Therefore the matrix
$(a_{ij}^{(k)})_{s\times (s-1)}$ has rank $s-1$. Without loss of generality, 
we assume that the last $(s-1)$ rows of the matrix are linearly independent. Then
the linear system
\begin{equation}  \label{h2}
\sum^{s}_{i=2}a_{ij}^{(k)}h_{i}=\alpha_{j}e^{\lambda_{j}\triangle_{k-1}}
-\sum^{n}_{i=s+1}a^{(k)}_{ij}h_{i}, 
\hspace{1cm} j=2, \dots, s,
\end{equation}
has a unique solution for $h_{i} \; (i=2, \dots, s)$. With $h_{i}\; (i=2, \dots, n)$
given by (\ref{h1}) and (\ref{h2}), the only matching condition left to be satisfied
now is 
\begin{equation} \label{cond4}
\sum^{s}_{i=2}a_{i1}^{(k)}h_{i}=\alpha_{1}e^{\lambda_{1}\triangle_{k-1}}
-\sum^{n}_{i=s+1}a^{(k)}_{i1}h_{i},
\end{equation}
which will determine the spacings of this $N$-pulse wavetrain. Since the matrix
$(a_{ij}^{(k)})_{s\times (n-1)}$ is not readily available, to determine the spacings
from Eq. (\ref{cond4}) is difficult. But this can be easily done with the aid of
the solution $\Psi^{(k)}$ of the adjoint equation (\ref{adjoint2}). 
With $h_{i}\; (i=2, \dots, n)$ given by (\ref{h1}) and (\ref{h2}), 
it is easy to show that Eqs. (\ref{phiasym1}) and (\ref{phiasym2})
become 
\begin{equation}
\tilde{\Phi}^{(k)}(x-x_{k}) \longrightarrow
\sum_{j=s+1}^{n}c_{j}^{(k+1)}\xi_{j}(x-x_{k+1})e^{\lambda_{j}(x-x_{k+1})}, 
\hspace{1cm} x_{k}\ll x\ll x_{k+1}, 
\end{equation}
and 
\begin{equation}
\tilde{\Phi}^{(k)}(x-x_{k}) \longrightarrow
w\; \xi_{1}(x-x_{k-1})e^{\lambda_{1}(x-x_{k-1})}+
\sum_{j=2}^{s}c_{j}^{(k-1)}\xi_{j}(x-x_{k-1})e^{\lambda_{j}(x-x_{k-1})},
\hspace{0.8cm} x_{k-1}\ll x\ll x_{k}, 
\end{equation}
where $w$ is a constant. Condition (\ref{cond4}) is equivalent to 
\begin{equation}  \label{cond5}
w=c_{1}^{(k-1)}.
\end{equation}
For $x_{k-1}\ll y_{1}\ll x_{k}$ and $x_{k}\ll y_{2}\ll x_{k+1}$, we have
\begin{equation}
\begin{array}{rcl}
0&=&\int^{y_{2}}_{y_{1}}\{d\tilde{\Phi}^{(k)}(x-x_{k})/dx-
\nabla G(\Phi^{(k)}(x-x_{k}))\tilde{\Phi}^{(k)}(x-x_{k})\}\cdot 
\Psi^{(k)*}(x-x_{k})dx \\
&=&\tilde{\Phi}^{(k)}(x-x_{k}))\cdot \Psi^{(k)*}(x-x_{k}) |^{y_{2}}_{y_{1}} \\
&&+\int^{y_{2}}_{y_{1}}\tilde{\Phi}^{(k)}(x-x_{k})\cdot \{-d\Psi^{(k)}(x-x_{k})/dx-
\nabla G^{*\mbox{\tiny\it T}}
(\Phi^{(k)}(x-x_{k}))\Psi^{(k)}(x-x_{k})\}^{*}dx \\
&=&\tilde{\Phi}^{(k)}(x-x_{k}))\cdot \Psi^{(k)*}(x-x_{k}) |^{y_{2}}_{y_{1}}.
\end{array}
\end{equation}
For $\triangle_{k-1}\gg 1$ and $\triangle_{k}\gg 1$, asymptotically we get
\begin{equation}
w\: d_{1}^{(k)*}e^{\lambda_{1}\triangle_{k-1}}+
\sum^{s}_{j=2}c_{j}^{(k-1)}d_{j}^{(k)*}e^{\lambda_{j}\triangle_{k-1}}
=\sum^{n}_{j=s+1}c_{j}^{(k+1)}d_{j}^{(k)*}e^{-\lambda_{j}\triangle_{k}}.
\end{equation}
Condition (\ref{cond5}) is satisfied if and only if Eq. (\ref{formula1}b) is valid. 
For the first and last waves in this $N$-pulse wavetrain, the analysis is simpler, 
and we get Eqs. (\ref{formula1}a,c) for matching. 
In summary, the $N$ pulses $\{\Phi^{(1)}, \dots, \Phi^{(N)}\}$ can be matched
and form a $N$-pulse wavetrain if and only if the spacings $\triangle_{k} (\gg 1)\;
(k=1, \dots, N-1)$ asymptotically satisfy the $N$ conditions (\ref{formula1}). 

Now we discuss the accuracy of the above results. Error is created mainly by
the matching requirements (see Eqs. (\ref{cond1}) and (\ref{cond2})) and the negligence
of nonlinear terms in Eq. (\ref{linear1}). 
First we discuss the error in the matching requirements. Let us reconsider the
solution (\ref{phi}) around the $k$-th wave. When $x_{k}\ll x\ll x_{k+1}$, beside
the exponentially decaying terms in $\Phi^{(k)}(x-x_{k})$, there are also such
terms in $\tilde{\Phi}^{(k)}(x-x_{k})$ (see Eq. (\ref{solution})). 
The combined exponentially decaying tails are
\begin{equation}
c_{1}^{(k)}\xi_{1}(x-x_{k})e^{\lambda_{1}(x-x_{k})}
+\sum^{s}_{j=2}(c_{j}^{(k)}+h_{j})\xi_{j}(x-x_{k})e^{\lambda_{j}(x-x_{k})}.
\end{equation}
Thus Eqs. (\ref{formula1}) would be more accurate if 
the $c_{j}^{(k)}$ values are replaced by $c_{j}^{(k)}+h_{j}$. Recall that 
$h_{j}\; (j=2, \dots, s)$ are determined from Eq. (\ref{h2}), so they are exponentially
small for large $\triangle_{k-1}$ and $\triangle_{k}$. 
As a result, the negligence of tail contribution 
from $\tilde{\Phi}^{(k)}(x-x_{k})$ only causes exponentially small relative errors
in Eqs. (\ref{formula1}). Simple reasoning also shows that
the exclusion of nonlinear terms in Eq. (\ref{linear}) also causes only 
exponentially small relative errors in (\ref{formula1}). 
The proof of theorem 1 is now completed. It should be pointed out that, 
if the eigenvalues \{$\lambda_{i},\; i=1, \dots, s$\} or
\{$\lambda_{i},\; i=s+1, \dots, n$\} are real-valued and close to each other, 
those exponentially small 
relative errors in Eqs. (\ref{formula1}) may become significant. 
In such cases, caution is needed in interpreting the results from (\ref{formula1})

\subsection{General permanent-wave trains}
The results in the previous section can be readily extended to the matching
of permanent waves which exponentially approach a complex constant at infinity. Suppose
such permanent waves exist in Eq. (\ref{ode}), then we make the following general
assumptions: for any permanent wave $\Phi(x)$ where
\begin{equation}
\Phi(x) \longrightarrow
\left\{ \begin{array}{l}
b_{2}, \hspace{0.5cm} x \rightarrow \infty, \\
b_{1}, \hspace{0.5cm} x \rightarrow -\infty, 
\end{array} \right.
\end{equation}
\begin{enumerate}
\item[B1.] the eigenvalues of the constant (Jacobian) matrices $\nabla G(b_{1})$ 
and $\nabla G(b_{2})$ all have non-zero real parts, and the number of $\nabla G(b_{1})$'s 
eigenvalues with negative real parts is equal to that of $\nabla G(b_{2})$'s 
eigenvalues with negative real parts; 
\item[B2.] the linear behavior dominates at infinity, i.e., as $x \rightarrow -\infty$
and $\infty$, $\Phi(x)$ approaches a solution of the linear equations 
\begin{equation} 
d\tilde{\Phi}/dx=\nabla G(b_{1})\tilde{\Phi},
\end{equation}
and 
\begin{equation} 
d\tilde{\Phi}/dx=\nabla G(b_{2})\tilde{\Phi},
\end{equation}
respectively; 
\item[B3.] the linearized equation of (\ref{ode}) around $\Phi(x)$
\begin{equation}  \label{linear2}
d\tilde{\Phi}/dx=\nabla G(\Phi)\tilde{\Phi}
\end{equation}
and its adjoint equation
\begin{equation}  \label{adjoint3}
-d\Psi/dx=\nabla G^{*\mbox{\tiny\it T}}(\Phi)\Psi
\end{equation}
each have a single linearly independent localized solution.
\end{enumerate}
Now suppose $\{\Phi^{(1)}, \Phi^{(2)}, \dots, \Phi^{(N)}\}$ are $N$ permanent waves with
\begin{equation}
\Phi^{(k)}(x) \longrightarrow
\left\{ \begin{array}{l}
b^{(k)}_{2}, \hspace{0.5cm} x \rightarrow \infty, \\
b^{(k)}_{1}, \hspace{0.5cm} x \rightarrow -\infty, 
\end{array} \right.
\end{equation}
where $1\le k\le N$. If they are to be matched and form a widely-separated
$N$-permanent-wave train, we need to require that 
\begin{equation} \label{match}
b_{1}^{(k)}=b_{2}^{(k-1)}, \hspace{0.3cm} 2\le k\le N. 
\end{equation}
One consequence is that all the matrices $\nabla G(b^{(k)}_{1})$ and $\nabla G(b^{(k)}_{2})$
$(1\le k\le N)$ have the same number of eigenvalues with negative real parts, which
we denote as $s$. We introduce the following notations. 
Denote $\nabla G(b^{(k)}_{1})$'s $n$ eigenvalues as 
$\lambda^{(k)}_{1}, \lambda^{(k)}_{2}, \dots, \lambda^{(k)}_{n}$ with
\begin{equation}
\mbox{Re}(\lambda^{(k)}_{1})\le \mbox{Re}(\lambda^{(k)}_{2})\le \dots \le \mbox{Re}(\lambda^{(k)}_{s})<0<
\mbox{Re}(\lambda^{(k)}_{s+1})\le \mbox{Re}(\lambda^{(k)}_{s+2})\le \dots \le \mbox{Re}(\lambda^{(k)}_{n}),
\end{equation}
and $\nabla G(b^{(k)}_{2})$'s as 
$\Lambda^{(k)}_{1}, \Lambda^{(k)}_{2}, \dots, \Lambda^{(k)}_{n}$ with
\begin{equation}
\mbox{Re}(\Lambda^{(k)}_{1})\le \mbox{Re}(\Lambda^{(k)}_{2})\le \dots \le \mbox{Re}(\Lambda^{(k)}_{s})<0<
\mbox{Re}(\Lambda^{(k)}_{s+1})\le \mbox{Re}(\Lambda^{(k)}_{s+2})\le \dots \le \mbox{Re}(\Lambda^{(k)}_{n}).
\end{equation}
The fundamental sets of solutions of the linear equations
\begin{equation} \label{lineark1}
d\tilde{\Phi}/dx=\nabla G(b^{(k)}_{1})\tilde{\Phi}
\end{equation}
and 
\begin{equation} \label{lineark2}
d\tilde{\Phi}/dx=\nabla G(b^{(k)}_{2})\tilde{\Phi}
\end{equation}
are respectively
$\{\xi^{(k)}_{i}(x)e^{\lambda_{i}^{(k)} x}, \; i=1,\dots, n\}$
and $\{\theta^{(k)}_{i}(x)e^{\Lambda_{i}^{(k)} x}, \; i=1,\dots, n\}$
which consist of chains of linearly independent solutions of Eqs. (\ref{lineark1})
and (\ref{lineark2}) as defined before (see Eq. (\ref{ksix})). The fundamental 
matrices of the adjoint equations of (\ref{lineark1}) and  (\ref{lineark2})
are then 
\begin{equation}
[\eta^{(k)}_{1}e^{-\lambda^{(k)*}_{1}x} \;\; 
\eta_{2}^{(k)}e^{-\lambda^{(k)*}_{2}x}\; \dots \;\;
\eta_{n}^{(k)}e^{-\lambda^{(k)*}_{n}x}] 
\end{equation}
and 
\begin{equation}
[\zeta^{(k)}_{1}e^{-\Lambda^{(k)*}_{1}x} \;\;
\zeta^{(k)}_{2}e^{-\Lambda^{(k)*}_{2}x}\;\; \dots \;\;
\zeta^{(k)}_{n}e^{-\Lambda^{(k)*}_{n}x}],
\end{equation}
with 
\begin{equation}
[\eta^{(k)}_{1}\;\; \eta_{2}^{(k)} \;\; \dots \;\; \eta_{n}^{(k)}]
=\{[\xi^{(k)}_{1}\;\; \xi^{(k)}_{2} \;\; \dots \;\; 
\xi^{(k)}_{n}]^{-1}\}^{*\mbox{\tiny\it T}}
\end{equation}
and
\begin{equation}
[\zeta^{(k)}_{1}\;\; \zeta^{(k)}_{2}\;\; \dots \;\; \zeta^{(k)}_{n}]
=\{[\theta^{(k)}_{1}\;\; \theta^{(k)}_{2} \;\;\dots \;\; 
\theta^{(k)}_{n}]^{-1}\}^{*\mbox{\tiny\it T}}. 
\end{equation}
Note that 
\begin{equation}
\lambda_{i}^{(k)}=\Lambda_{i}^{(k-1)}, \;\;
\xi_{i}^{(k)}=\theta_{i}^{(k-1)}, \;\;
\eta_{i}^{(k)}=\zeta_{i}^{(k-1)}, \hspace{0.5cm} i=1, \dots, n 
\end{equation}
in view of (\ref{match}). In those notations, we have 
\begin{equation} 
\Phi^{(k)}(x) \longrightarrow
\left\{ \begin{array}{l}
b_{2}^{(k)}+\sum_{i=1}^{s}c_{i}^{(k)}
\theta_{i}^{(k)}(x)e^{\Lambda^{(k)}_{i}x}, \hspace{0.6cm} x\rightarrow \infty, \\
b_{1}^{(k)}+\sum_{i=s+1}^{n}c_{i}^{(k)}
\xi^{(k)}_{i}(x)e^{\lambda^{(k)}_{i}x}, \hspace{0.3cm} x\rightarrow -\infty,
\end{array} \right.
\end{equation}
according to assumption B2. 
For the single linearly independent localized solution $\Psi^{(k)}$ of the 
adjoint equation
\begin{equation}  \label{adjoint4}
-d\Psi^{(k)}/dx=\nabla G^{*\mbox{\tiny\it T}}(\Phi^{(k)})\Psi^{(k)},
\end{equation}
we have 
\begin{equation}
\Psi^{(k)}(x) \longrightarrow
\left\{ \begin{array}{l}
\sum^{s}_{i=1}d^{(k)}_{i}\eta^{(k)}_{i}(x)e^{-\lambda^{(k)*}_{i}x},
\hspace{1.5cm} x\rightarrow -\infty, \\
\sum^{n}_{i=s+1}d^{(k)}_{i}\zeta^{(k)}_{i}(x)e^{-\Lambda^{(k)*}_{i}x}, 
\hspace{1.1cm} x\rightarrow \infty.
\end{array} \right.
\end{equation}
Here $c_{i}^{(k)}$ and $d_{i}^{(k)}$ $(1\le i\le n)$ are complex constants. 

Now consider a widely-separated permanent-wave train matched by the above
$N$ permanent waves $\{\Phi^{(1)}$, $\Phi^{(2)}$, \dots, $\Phi^{(N)}\}$. Assume that
the $k$-th wave $\Phi^{(k)}$ is located at $x=x_{k}$ $(k=1, 2, \dots, N)$, and
$\triangle_{k}$ is as defined in Eq. (\ref{deltak}), then we have the following 
result. 

\begin{guess}  
Under the assumptions B1, B2, B3, (\ref{match})
and the above notations, the $N$ permanent waves
$\{\Phi^{(1)}, \dots, \Phi^{(N)}\}$ can match each other and form a widely-separated
$N$-permanent-wave train if and only if the spacings 
$\triangle_{k} \;(\gg 1)\; (k=1, \dots, N-1)$
asymptotically satisfy the following $N$ conditions
\begin{subequations} \label{formula2}
\begin{equation}
\sum^{n}_{j=s+1}c_{j}^{(2)}d_{j}^{(1)*}e^{-\Lambda^{(1)}_{j}\triangle_{1}}=0,
\end{equation}
\begin{equation}
\sum^{s}_{j=1}c_{j}^{(k-1)}d_{j}^{(k)*}e^{\lambda^{(k)}_{j}\triangle_{k-1}}
=\sum^{n}_{j=s+1}c_{j}^{(k+1)}d_{j}^{(k)*}e^{-\Lambda^{(k)}_{j}\triangle_{k}},
\hspace{0.5cm} (2\le k\le N-1),
\end{equation}
\begin{equation}
\sum^{s}_{j=1}c_{j}^{(N-1)}d_{j}^{(N)*}e^{\lambda^{(N)}_{j}\triangle_{N-1}}=0.
\end{equation}
\end{subequations}
The relative errors in Eqs. (\ref{formula2}) are exponentially small with the spacings.
\end{guess}
The proof for this theorem is similar to that for theorem 1, and is thus omitted here. 

Remark: In applying theorem 2 to a given nonlinear wave system, the 
major difficulty is the determination of the coefficients $\{d_{j}^{(k)}\}$
in the localized solution $\Psi^{(k)}(x)$ of the adjoint equation (\ref{adjoint4}). 
In general, this has to be done numerically. But in many cases, 
Eq. (\ref{linear2}) can 
be cast into a self-adjoint system (see \cite{yang}, \cite{buffoni} and 
\cite{vandervorst}). Then $\Psi^{(k)}(x)$ and its coefficients $\{d_{j}^{(k)}\}$
can be readily obtained from $d\Phi^{(k)}/dx$, and the verification of conditions
(\ref{formula2}) can proceed. 

\section{Applications}
\subsection{Fourth-order systems}
The permanent waves in many nonlinear wave problems are governed by fourth-order
systems (\ref{ode}) (see \cite{yang}, \cite{buffoni} and
\cite{vandervorst}). In this section, we apply theorems 1 and 2 to certain classes of
such systems. In particular, we will establish the existence of countably infinite 
multiple permanent-wave trains under some general assumptions. 

We first consider the matching of identical permanent waves in a 
fourth-order system (\ref{ode}). Suppose $\Phi(x)$ is a permanent wave in
(\ref{ode}) where $\Phi(x) \rightarrow b$ as $|x|\rightarrow \infty$, and the
assumptions B1, B2 and B3 are satisfied. Moreover, we suppose 
the eigenvalues of $\nabla G(b)$ are $\pm \lambda_{1}$ and $\pm \lambda_{2}$, 
where $\lambda_{1}\ne \lambda_{2}$ and $\mbox{Re}(\lambda_{i})>0\; (i=1, 2)$. 
Then corresponding to the four 
distinct eigenvalues $-\lambda_{1}, -\lambda_{2}, \lambda_{1}$ and $\lambda_{2}$,
$\nabla G(b)$ has four linearly independent eigenvectors $\xi_{i} \;(i=1, \dots, 4)$. 
If we denote 
\begin{equation}  \label{eta}
[\eta_{1}\; \eta_{2} \; \eta_{3} \; \eta_{4}]
=\{[\xi_{1}\; \xi_{2} \; \xi_{3} \; \xi_{4}]^{-1}\}^{*\mbox{\tiny\it T}}, 
\end{equation}
then 
\begin{equation}
\Phi(x) \longrightarrow 
\left\{ \begin{array}{l} 
b+c_{1}\xi_{1}e^{-\lambda_{1}x}+c_{2}\xi_{2}e^{-\lambda_{2}x}, \hspace{1cm}
 x \rightarrow \infty, \\
b+c_{3}\xi_{3}e^{\lambda_{1}x}+c_{4}\xi_{4}e^{\lambda_{2}x}, \hspace{1.5cm}
 x \rightarrow -\infty, 
\end{array} \right.
\end{equation}
and
\begin{equation}
\Psi(x) \longrightarrow 
\left\{ \begin{array}{l}  
d_{1}\eta_{1}e^{\lambda^{*}_{1}x}+d_{2}\eta_{2}e^{\lambda^{*}_{2}x}, \hspace{1.5cm}
 x \rightarrow -\infty, \\
d_{3}\eta_{3}e^{-\lambda^{*}_{1}x}+d_{4}\eta_{4}e^{-\lambda^{*}_{2}x}, \hspace{1cm}
 x \rightarrow \infty. 
\end{array} \right.
\end{equation}
For some fourth-order problems, Eq. (\ref{linear2}) can be cast into
a self-adjoint system, and one has either $\lambda_{1}$ and $\lambda_{2}$ real-valued
with $(d_{1}\; d_{3}) \propto (c_{3}\; c_{1})$ and
$(d_{2}\; d_{4}) \propto (c_{4}\; c_{2})$, or
$\lambda_{1}$ and $\lambda_{2}$ complex-valued with $\lambda_{2}=\lambda^{*}_{1}$, 
$(d_{1}\; d_{3}) \propto (c_{4}\; c_{2})$ and
$(d_{2}\; d_{4}) \propto (c_{3}\; c_{1})$. 
In such cases, conditions
(\ref{formula2}) for the matching of $N$ identical permanent waves 
$\{\Phi(x), \dots, \Phi(x)\}$ simply become 
\begin{equation} \label{cond6}
c_{1}d^{*}_{1}e^{-\lambda_{1}\triangle_{k}}+
c_{2}d^{*}_{2}e^{-\lambda_{2}\triangle_{k}}=0, \hspace{0.5cm}
k=1, \dots, N-1.
\end{equation}
In the second case, if furthermore (\ref{ode}) is a real system, then 
$\lambda_{2}=\lambda_{1}^{*}$, 
$c_{2}=c_{1}^{*}, \hspace{0.1cm} d_{2}=d_{1}^{*}$, and Eq. (\ref{cond6})
becomes 
\begin{equation} \label{cond7}
c_{1}d^{*}_{1}e^{-i\: \mbox{Im}(\lambda_{1})\: \triangle_{k}}+
c_{1}^{*}d_{1}e^{i\: \mbox{Im}(\lambda_{1})\: \triangle_{k}}=0, \hspace{0.5cm} 
k=1, \dots, N-1.
\end{equation}
The spacings $\triangle_{k}$ can then be easily obtained from (\ref{cond7}) as
\begin{equation} \label{spacing}
\triangle_{k}=(\mbox{arg}(c_{1}d_{1}^{*})
+\frac{\pi}{2}+m_{k}\pi)/\mbox{Im}(\lambda_{1}), 
\hspace{0.5cm} k=1, \dots, N-1,
\end{equation}
where $m_{k}$ is any non-negative integer. 
Note that in this case, the exponentially small relative
errors in (\ref{formula2}) make little difference, especially when $m_{k}$ is large. 
Thus we conclude that an arbitrary number of identical permanent waves $\Phi(x)$ 
can be matched together and form multiple permanent-wave trains, whose spacings
are given asymptotically by Eq. (\ref{spacing}). Clearly a countably infinite 
number of such wavetrains can be formed. In the paper by Buffoni and Sere 
\cite{buffoni}, they proved the existence of countably infinite multi-pulse permanent
wave solutions for a class of coupled-nonlinear-Schr\"odinger-type equations. 
When those equations are cast into a fourth-order system of the two variables and 
their first derivatives, it is easy to check that they fall into the above category. 
Thus their result is a special case of ours. 
But differences also exist between their result and ours. In their result, 
$m_{k}$ in Eq. (\ref{spacing}) 
is an even integer; while in ours, it is any integer. This means that
we identified twice as many solitary-wave trains as they did. 
For fourth-order systems where $\Phi(x)-b$ and  $\Psi(x)$ 
element-wise are both even or odd in $x$, or one of them is even (odd) and the other 
one odd (even), then 
$(c_{3}\; c_{4})=\pm (c_{1}\; c_{2})$, and
$(\xi_{3}\; \xi_{4})$ is row-wise equal to or opposite of $(\xi_{1}\; \xi_{2})$. 
It is easy to show from (\ref{eta}) that $(\eta_{3}\; \eta_{4})$ is 
also row-wise equal to or opposite of $(\eta_{1}\; \eta_{2})$
and  $(d_{3}\; d_{4})=\pm (d_{1}\; d_{2})$. 
Thus the $N$ matching conditions (\ref{formula2}) also reduce to (\ref{cond6}). 
If further more, $\lambda_{2}=\lambda_{1}^{*}$, then we will find 
countably infinite multiple permanent-wave trains whose spacings are given 
by (\ref{spacing}). 

Next we consider the matching of different permanent waves in a fourth-order
system (\ref{ode}). Suppose $G$ is an odd function of $\Phi$, i.e., 
\begin{equation} \label{odd}
G(-\Phi)=-G(\Phi), 
\end{equation}
and $\Phi(x)$ is a permanent wave in Eq. (\ref{ode}) with
\begin{equation}
\Phi(x) \longrightarrow
\left\{ \begin{array}{l}
b, \hspace{0.8cm} x \rightarrow \infty, \\
-b, \hspace{0.5cm} x \rightarrow -\infty, 
\end{array} \right.
\end{equation}
then $-\Phi(x)$ is also a permanent wave in (\ref{ode}). It is easy to show
from (\ref{odd}) that $\nabla G(-\Phi)=\nabla G(\Phi)$, 
thus $\nabla G(-b)=\nabla G(b)$. 
Beside the assumptions B1, B2 and B3, we also assume that $\nabla G(b)$'s four
eigenvalues are $\pm \lambda_{1}$ and $\pm \lambda_{2}$ with
$\lambda_{1}\ne \lambda_{2}$ and $\mbox{Re}(\lambda_{i})>0\; (i=1, 2)$.
Suppose the eigenvectors corresponding to $-\lambda_{1}, -\lambda_{2}, \lambda_{1}$ 
and $\lambda_{2}$ are denoted as $\xi_{i} (i=1, \dots, 4)$, then we have
\begin{equation}
\Phi(x) \longrightarrow 
\left\{ \begin{array}{l} 
b+c_{1}\xi_{1}e^{-\lambda_{1}x}+c_{2}\xi_{2}e^{-\lambda_{2}x}, \hspace{1cm}
 x \rightarrow \infty, \\
-b+c_{3}\xi_{3}e^{\lambda_{1}x}+c_{4}\xi_{4}e^{\lambda_{2}x}, \hspace{1.15cm}
 x \rightarrow -\infty.
\end{array} \right.
\end{equation}
For the localized solution $\Psi(x)$ of the adjoint equation
(\ref{adjoint3}), we have
\begin{equation}
\Psi(x) \longrightarrow 
\left\{ \begin{array}{l}  
d_{1}\eta_{1}e^{\lambda^{*}_{1}x}+d_{2}\eta_{2}e^{\lambda^{*}_{2}x}, \hspace{1.4cm}
 x \rightarrow -\infty, \\
d_{3}\eta_{3}e^{-\lambda^{*}_{1}x}+d_{4}\eta_{4}e^{-\lambda^{*}_{2}x}, \hspace{0.9cm}
 x \rightarrow \infty,
\end{array} \right.
\end{equation}
where $\eta_{i}\; (i=1, \dots, 4)$ are given by (\ref{eta}). For those equations
(\ref{ode}) where Eq. (\ref{linear2}) can be cast into
a self-adjoint system and one has either
$(d_{1}\; d_{3}) \propto (c_{3}\; c_{1})$ and
$(d_{2}\; d_{4}) \propto (c_{4}\; c_{2})$ with $\lambda_{1}$ and $\lambda_{2}$ real, 
or $(d_{1}\; d_{3}) \propto (c_{4}\; c_{2})$ and
$(d_{2}\; d_{4}) \propto (c_{3}\; c_{1})$ with $\lambda_{2}=\lambda^{*}_{1}$, 
conditions (\ref{formula2}) for the matching of permanent waves $\{\Phi, 
-\Phi, \Phi, -\Phi, \dots\}$ or $\{-\Phi, \Phi, -\Phi, \Phi, \dots\}$
will also reduce to (\ref{cond6}). When $\lambda_{2}=\lambda^{*}_{1}$, if 
furthermore (\ref{ode}) is a real system, then 
we can show as before that such matchings are always possible and the
spacings are given by Eq. (\ref{spacing}). An infinite number of such wavetrains
will be obtained. We point out that the fourth-order systems studied by Kalies
and VanderVorst \cite{vandervorst} falls into this category and is thus a 
special case of the above results. Here again we identified twice as many 
permanent-wave trains as they did since $m_{k}$ in Eq. (\ref{spacing})
needs to be an even integer in their result. 

\subsection{Coupled nonlinear Schr\"odinger equations}
The coupled nonlinear Schr\"odinger equations govern the evolution of two
interacting wave packets in nonlinear and dispersive physical systems
\cite{benney}. These equations are particularly important in nonlinear optics
as they govern the pulse propagation in birefringent nonlinear optical fibers
\cite{menyuk}. In recent years, the experimental design of high-speed
optical-soliton-based telecommunication systems stimulated great interest
in these equations, and much work has been done on them (see \cite{agrawal}
and the references therein). In particular, simple and multi-pulse
solitary waves in these equations 
have been found and classified in \cite{yang}. In this section, 
we study the multiple permanent-wave trains in these equations. 
We primarily discuss the focusing case where solitary waves exist. In the end of 
this section, we comment on the defocusing case where dark solitons arise. 

The solitary waves in coupled nonlinear Schr\"odinger equations (focusing case)
are governed by the following set of equations 
\begin{subequations} \label{r1r2}
\begin{equation}
r_{1xx}-r_{1}+(r_{1}^{2}+\beta r_{2}^{2})r_{1}=0,
\end{equation}
\begin{equation}
r_{2xx}-\omega^{2}r_{2}+(r_{2}^{2}+\beta r_{1}^{2})r_{2}=0,
\end{equation}
\end{subequations}
where $r_{1}$ and $r_{2}$ approach zero as $x$ goes to infinity, and 
$\beta$ and $\omega$ are positive parameters. 
To apply theorem 1 to these equations, we first rewrite them as the following
first order system
\begin{equation} \label{U}
dU/dx=G(U),
\end{equation}
where 
\begin{equation}
U=(u_{1}, u_{2}, u_{3}, u_{4})^{T}=(r_{1}, r_{1x}, r_{2}, r_{2x})^{T},
\end{equation}
and 
\begin{equation}
G(U)=\left(\begin{array}{c}
u_{2} \\ u_{1}-(u_{1}^{2}+\beta u_{3}^{2})u_{1} \\
u_{4} \\ \omega^{2}u_{3}-(u_{3}^{2}+\beta u_{1}^{2})u_{3}
\end{array} \right).
\end{equation}
It is easy to check that the above system satisfies the assumptions A1, A2 and A3
when $\beta\ne 1$. Thus in the following we assume that $\beta\ne 1$. 
The eigenvalues of the matrix $\nabla G(0)$ are $-1, -\omega, 1$ and $\omega$, and the
corresponding eigenvectors are 
\begin{equation}
\xi_{1}=\left(\begin{array}{c}1 \\ -1 \\ 0 \\ 0 \end{array}\right),  \;\;
\xi_{2}=\left(\begin{array}{c}0 \\ 0 \\ 1 \\ -\omega \end{array}\right), \;\;
\xi_{3}=\left(\begin{array}{c}1 \\ 1 \\ 0 \\ 0 \end{array}\right), \;\;
\xi_{2}=\left(\begin{array}{c}0 \\ 0 \\ 1 \\ \omega \end{array}\right).
\end{equation}
For a solitary wave $(r_{1}, r_{2})$ of Eqs. (\ref{r1r2}) with
\begin{equation}
r_{1}(x) \longrightarrow \left\{ \begin{array}{l}
c_{1}e^{-x}, \hspace{0.8cm} x \rightarrow \infty, \\
c_{3}e^{x}, \hspace{1cm} x \rightarrow -\infty, 
\end{array} \right.
\end{equation}
and
\begin{equation}
r_{2}(x) \longrightarrow \left\{ \begin{array}{l}
c_{2}e^{-\omega x}, \hspace{0.8cm} x \rightarrow \infty, \\
c_{4}e^{\omega x}, \hspace{1cm} x \rightarrow -\infty, 
\end{array} \right.
\end{equation}
we have
\begin{equation}
U(x) \longrightarrow \left\{ \begin{array}{l}
c_{1}\xi_{1}e^{-x}+c_{2}\xi_{2}e^{-\omega x}, \hspace{1cm} x \rightarrow \infty,\\
c_{3}\xi_{3}e^{x}+c_{4}\xi_{4}e^{\omega x}, \hspace{1.5cm} x \rightarrow -\infty.
\end{array} \right.
\end{equation}
The linearized equation of (\ref{U}) around a solitary wave $U(x)$ is 
\begin{equation}  \label{tildeU}
d\tilde{U}/dx=\nabla G(U) \tilde{U}, 
\end{equation}
where $\tilde{U}=(\tilde{u}_{1}, \tilde{u}_{2}, \tilde{u}_{3}, \tilde{u}_{4})^{T}$, 
and
\begin{equation}
\nabla G(U)=\left(\begin{array}{cccc}
0 & 1 & 0 & 0 \\
1-3u_{1}^{2}-\beta u_{3}^{2} & 0 &-2\beta u_{1} u_{3} & 0 \\
0 & 0 & 0 & 1 \\
-2\beta u_{1} u_{3} & 0 & 1-3u_{3}^{2}-\beta u_{1}^{2} & 0
\end{array} \right).
\end{equation}
The single localized solution of the above equations is $dU/dx$. If $\tilde{u}_{2}$ 
and $\tilde{u}_{4}$ in (\ref{tildeU}) are eliminated in favor of $\tilde{u}_{1}$
and $\tilde{u}_{3}$, then the linear system for $\tilde{u}_{1}$ and $\tilde{u}_{3}$
are self-adjoint. The adjoint equation of (\ref{tildeU}) is
\begin{equation} \label{V}
-dV/dx=\nabla G^{\mbox{\tiny\it T}}(U)V
\end{equation}
with $V=(v_{1}, v_{2}, v_{3}, v_{4})$. It is easy to see that when $v_{1}$ and
$v_{3}$ are eliminated from  (\ref{V}), the equations for 
$v_{2}$ and $v_{4}$ are the same as those for $\tilde{u}_{1}$ and $\tilde{u}_{3}$.
Thus the single localized solution of Eqs. (\ref{V}) is 
\begin{equation}
V=(-u_{1xx}, u_{1x}, -u_{3xx}, u_{3x})^{T}. 
\end{equation}
At infinity, 
\begin{equation}
V(x) \longrightarrow \left\{ \begin{array}{l}
d_{1}\eta_{1}e^{x}+d_{2}\eta_{2}e^{\omega x}, \hspace{1.5cm} x \rightarrow -\infty,\\
d_{3}\eta_{3}e^{-x}+d_{4}\eta_{4}e^{-\omega x}, \hspace{1cm} x \rightarrow \infty,
\end{array} \right.
\end{equation}
where $\{\eta_{i}, i=1, \dots, 4\}$ are obtained from Eq. (\ref{eta}) as
\begin{equation}
\eta_{1}=\left(\begin{array}{c}\frac{1}{2} \\ -\frac{1}{2} 
\\ 0 \\ 0 \end{array}\right),  \;\;
\eta_{2}=\left(\begin{array}{c}0 \\ 0 \\ \frac{1}{2} 
\\ -\frac{1}{2\omega} \end{array}\right), \;\;
\eta_{3}=\left(\begin{array}{c}\frac{1}{2} \\ \frac{1}{2} 
\\ 0 \\ 0 \end{array}\right),  \;\;
\eta_{4}=\left(\begin{array}{c}0 \\ 0 \\ \frac{1}{2} 
\\ \frac{1}{2\omega} \end{array}\right),
\end{equation}
and
\begin{equation} \label{d}
d_{1}=-2c_{3}, \;\; d_{3}=-2c_{1}, \;\; d_{2}=-2\omega^{2}c_{4}, \;\;
d_{4}=-2\omega^{2}c_{2}.
\end{equation}
Now we consider the matching of $N$ solitary waves $\{(r_{1}^{(k)}, r_{2}^{(k)}), 
k=1, \dots, N\}$ where 
\begin{equation}
r_{1}^{(k)}(x) \longrightarrow \left\{ \begin{array}{l}
c_{1}^{(k)}e^{-x}, \hspace{0.8cm} x \rightarrow \infty, \\
c_{3}^{(k)}e^{x}, \hspace{1cm} x \rightarrow -\infty, 
\end{array} \right.
\end{equation}
and
\begin{equation}
r_{2}^{(k)}(x) \longrightarrow \left\{ \begin{array}{l}
c_{2}^{(k)}e^{-\omega x}, \hspace{0.8cm} x \rightarrow \infty, \\
c_{4}^{(k)}e^{\omega x}, \hspace{1cm} x \rightarrow -\infty.
\end{array} \right.
\end{equation}
In view of (\ref{d}), the matching conditions (\ref{formula1}) become
\begin{equation} \label{cond8}
c_{1}^{(k)}c_{3}^{(k+1)}e^{-\triangle_{k}}+\omega^{2}c_{2}^{(k)}
c_{4}^{(k+1)}e^{-\omega \triangle_{k}}=0, \hspace{0.5cm} k=1, \dots, N-1.
\end{equation}
Interestingly Eq. (\ref{cond8}) indicates that the $N$ solitary waves 
$\{(r_{1}^{(k)}, r_{2}^{(k)}),\; k=1, \dots, N\}$ can be matched if and 
only if all the adjacent solitary waves can. Thus the matching of solitary waves
in Eqs. (\ref{r1r2}) is a ``local'' phenomenon. This fact would make the
construction of those multiple solitary-wave trains much easier. 
In what follows, we discuss the matching of some special types of solitary waves. 

First we consider the matching of wave and daughter wave solutions. In such solutions, 
either $r_{2}\ll r_{1}$ or $r_{1}\ll r_{2}$. Without loss of generality, we
assume that $r_{2}\ll r_{1}$. These solutions 
exist near the curves 
\begin{equation} \label{boundary}
\omega=(\sqrt{1+8\beta}-1)/2-m
\end{equation}
in the $(\omega, \beta)$ parameter plane \cite{yang}. 
Here $m$ is a non-negative integer and $m < (\sqrt{1+8\beta}-1)/2$.
In these solutions, $r_{1}$ is symmetric; $r_{2}$ is symmetric
for even values of $m$ and anti-symmetric for odd values of $m$. 
Suppose $(\hat{r}_{1}, \hat{r}_{2})$ is such a solution, then 
\begin{subequations} \label{daughter}
\begin{equation}
\hat{r}_{1} \longrightarrow c_{1}e^{-|x|}, \hspace{1cm} |x| \rightarrow \infty, 
\end{equation}
and
\begin{equation}
\hat{r}_{2} \longrightarrow \left\{ \begin{array}{l}
c_{2}e^{-\omega x}, \hspace{1.4cm} x \rightarrow \infty, \\
(-1)^{m}c_{2}e^{\omega x}, \hspace{0.6cm} x \rightarrow -\infty.
\end{array} \right.
\end{equation}
\end{subequations}
Here $c_{2}\ll 1$. Notice that 
if $r_{i}(x)$ ($i$=1 or 2) is a solution of Eqs. (\ref{r1r2}), so is $-r_{i}(x)$.
Without loss of generality, we require that $c_{i}>0 \; (i=1,2)$. 
Now we consider $N$ wave and daughter wave solutions
$\{(r_{1}^{(k)}, r_{2}^{(k)}), k=1, \dots, N\}$ where 
\begin{equation}
r_{1}^{(k)}(x)=q_{1}^{(k)}\hat{r}_{1}(x), \;\;
r_{2}^{(k)}(x)=q_{2}^{(k)}\hat{r}_{2}(x), \hspace{0.5cm}
k=1, \dots, N,
\end{equation}
and $q_{i}^{(k)}=\pm 1 \;(i=1, 2)$. The matching condition (\ref{cond8})
for these solitary waves are simply
\begin{equation} \label{cond11}
q_{1}^{(k)}q_{1}^{(k+1)}c_{1}^{2}e^{-\triangle_{k}}
+(-1)^{m}q_{2}^{(k)}q_{2}^{(k+1)}\omega^{2}c_{2}^{2}e^{-\omega \triangle_{k}}=0, 
\hspace{0.5cm} k=1, \dots, N-1, 
\end{equation}
i.e. 
\begin{equation}
e^{-(1-\omega) \triangle_{k}}=(-1)^{m+1}\frac{q_{2}^{(k)}q_{2}^{(k+1)}}
{q_{1}^{(k)}q_{1}^{(k+1)}} \frac{\omega^{2}c_{2}^{2}}{c_{1}^{2}}, 
\hspace{0.5cm} k=1, \dots, N-1.
\end{equation}
For these conditions to be satisfied, we need to require that $\omega<1$ and
\begin{equation} \label{cond9}
(-1)^{m+1}q_{1}^{(k)}q_{1}^{(k+1)}q_{2}^{(k)}q_{2}^{(k+1)}=1, 
\hspace{0.5cm} k=1, \dots, N-1.
\end{equation}
Suppose $(q_{1}^{(k)}, q_{2}^{(k)})$ is fixed, then condition (\ref{cond9}) shows that
$(q_{1}^{(k+1)},\; q_{2}^{(k+1)})$ can take two sets of values. In other words, 
there are two possible types of matching. Thus these $N$ wave and daughter wave
solutions can form $2^{N}$ topologically distinct solitary-wave trains. Since $N$
is arbitrary, countably infinite multiple-pulse solitary waves will be formed.
The spacings between adjacent waves in those wavetrains are
\begin{equation} \label{cond10}
\triangle_{k}=\frac{\ln(\omega^{2}c_{2}^{2}/c_{1}^{2})}{\omega-1}, 
\hspace{0.5cm} k=1, \dots, N-1,
\end{equation}
which are the same throughout an entire wavetrain. As $\omega$ approaches 
the wave and daughter wave boundary $(\sqrt{1+8\beta}-1)/2-m$, 
$c_{1}$ approaches $2\sqrt{2}$, $c_{2}$ approaches 0, and thus $\triangle_{k}$ 
approaches infinity. The above theoretical results can be checked numerically. 
We first select $(\beta, \omega)$ to be (2/3, 0.85) which is close to the
curve (\ref{boundary}) with $m$ equal to zero. With these parameter values, 
it is easy to find numerically that $c_{1}$ and $c_{2}$ as in Eq. (\ref{daughter})
are equal to 2.6592 and 1.1744 respectively. 
Eqs. (\ref{cond9}) and (\ref{cond10}) then predict
that the two wave and daughter waves $(\hat{r}_{1}, \hat{r}_{2})$ and
$(-\hat{r}_{1}, \hat{r}_{2})$ can be matched with the spacing approximately equal to 
13.0635. This is indeed the case. Numerically we found this exact two-pulse
solitary wave and plotted it in Fig. 1a. The exact spacing (measured as the 
distance between the two extrema in $r_{1}$) is 13.064, which is very close to 
the theoretical prediction. Next we select $(\beta, \omega)$ to be
(2, 0.6) which is close to the curve (\ref{boundary}) with $m=1$. In this case, 
we numerically found that $c_{1}$ and $c_{2}$ in (\ref{daughter})
are equal to 3.0386 and 0.6041. Then we predict from (\ref{cond9}) and (\ref{cond10})
that $(\hat{r}_{1}, \hat{r}_{2})$ and itself can be matched with the spacing 
approximately equal to 10.6308. Indeed, that exact two-pulse solution was numerically
found and plotted in Fig. 1b. The exact spacing is 10.40, close to the predicted value. 
The predictions on other types of matchings were also verified with good accuracy. 
We point out that each multiple-pulse solitary wave will generate a family of solitary
waves as the parameter pair $(\omega, \beta)$ moves away from the curves
(\ref{boundary}). Therefore countably infinite families of solitary waves will
be generated near those curves. 

Next we discuss mixed matchings between wave and daughter wave solutions and 
other types of solitary waves. When $(\omega, \beta)$
is near the curve (\ref{boundary}) with $m=0$, 
beside the wave and daughter wave solutions, another type of solitary waves
(belonging to family $D_{2}$) also exist \cite{yang}. Suppose 
$(\hat{r}_{1}, \hat{r}_{2})$ is a wave and daughter wave solution whose 
large $x$ behavior is given by (\ref{daughter}) (with $m=0$), and 
$(\bar{r}_{1}, \bar{r}_{2})$ is a solitary wave with 
\begin{subequations}
\begin{equation}
\bar{r}_{1} \longrightarrow \alpha_{1}e^{-|x|}, \hspace{1cm} |x| \rightarrow \infty, 
\end{equation}
\begin{equation}
\bar{r}_{2} \longrightarrow \alpha_{2}\: \mbox{sgn}(x)\: e^{-\omega |x|}, 
\hspace{1cm} |x| \rightarrow \infty,
\end{equation}
\end{subequations}
and $\alpha_{i}>0\; (i=1, 2)$. Consider the mixed matching of the solitary 
waves $(q_{1}\hat{r}_{1}, q_{2}\hat{r}_{2})$ and 
$(q_{3}\bar{r}_{1}, q_{4}\bar{r}_{2})$ where $q_{i}\; (i=1, \dots, 4)$ are either
1 or $-1$. The matching condition is 
\begin{equation}
q_{1}q_{3}c_{1}\alpha_{1}e^{-\triangle}-
q_{2}q_{4}\omega^{2}c_{2}\alpha_{2}e^{-\omega\triangle}=0, 
\end{equation}
or
\begin{equation}
e^{-(1-\omega)\triangle}=\frac{q_{2}q_{4}}{q_{1}q_{3}} \frac{\omega^{2}c_{2}\alpha_{2}}
{c_{1}\alpha_{1}}
\end{equation}
where $\triangle$ is the spacing. 
This condition can be satisfied if and only if $\omega<1$ and 
the sign of $q_{1}q_{2}q_{3}q_{4}$ is equal to 1. 
As an example, we choose $(\beta, \omega)$ as (2/3, 0.78). Then it is easy to find that 
$c_{1}$, $c_{2}$, $\alpha_{1}$ and $\alpha_{2}$ are 2.7967, 0.5210, 
7.8105 and 8.4171 respectively. 
The above results predict that $(\hat{r}_{1}, \hat{r}_{2})$ 
and $(-\bar{r}_{1}, -\bar{r}_{2})$ can match each other and form a new two-pulse
solitary wave. This was verified numerically. The exact matched solution is 
plotted in Fig. 2 with the spacing 10.26, while the predicted value for the spacing 
is 9.5571. Mixed matching between many copies of $(\hat{r}_{1}, \hat{r}_{2})$ and 
$(\bar{r}_{1}, \bar{r}_{2})$ can be similarly analysed. Once again, countably
infinite multiple-pulse solitary waves will be formed by these mixed matchings. 

Lastly we discuss the matching of solitary waves near $\omega=1$. In this case, 
single-hump solitary waves with $r_{1}\approx r_{2}$ are present. 
Suppose $(\hat{r}_{1}, \hat{r}_{2})$ is such a solution with
\begin{subequations}
\begin{equation}
\hat{r}_{1} \longrightarrow c_{1}e^{-|x|}, \hspace{1cm} |x| \rightarrow \infty, 
\end{equation}
\begin{equation}
\hat{r}_{2} \longrightarrow c_{2} e^{-\omega |x|}, 
\hspace{1cm} |x| \rightarrow \infty,
\end{equation}
\end{subequations}
then $c_{1}\approx c_{2}$. If we consider the matching of these solitary waves
$\{(q_{1}^{(k)}\hat{r}_{1}, q_{2}^{(k)}\hat{r}_{2})\}$ where 
$q_{i}^{(k)}=\pm 1\; (i=1, 2)$, the matching condition would again be Eq. (\ref{cond11})
(with $m=0$). But here since $\nabla G(0)$'s eigenvalues 1 and $\omega$ are close,
the exponentially small relative errors in (\ref{formula1}) and (\ref{cond11})
may become important. Thus condition (\ref{cond11}) should be treated with
caution. For instance, when $(\beta, \omega)$ is (2, 0.99), 
we found $c_{1}$ and $c_{2}$ to be 1.6142 and 1.6355. In this case, 
$\omega^{2}c_{2}^{2}/c_{1}^{2}>1$. Thus according to (\ref{cond11}), 
$(\hat{r}_{1}, \hat{r}_{2})$ and $(\hat{r}_{1}, -\hat{r}_{2})$ can not be 
matched. But our numerical results show otherwise \cite{yang}.

Theorems 1 and 2 can also be used to study the matching of dark solitons which exist
in coupled nonlinear Schr\"odinger equations (defocusing case) \cite{agrawal}. 
In this case, 
our results on the matching of some classes of dark solitons indicate that
such matchings are impossible since conditions (\ref{formula2}) can not be satisfied. 
We suspect that any dark solitons can not match each other to form widely-separated
dark-soliton trains. 

\section{Discussion}
The results in this paper can be readily applied to general nonlinear wave systems
for the construction of widely-separated multiple permanent-wave trains. 
Such wavetrains geometrically look like a superposition of individual permanent
waves. This is somewhat analogous to the superposition principle of solutions 
in a linear system. But the difference here is that, due to the nonlinear nature
of Eq. (\ref{ode}), those individual permanent waves have to be properly spaced
(according to Eq. (\ref{formula1}) or (\ref{formula2})) in order to form a 
wavetrain. When such wavetrains exist, one important question is
their stability. For the coupled nonlinear Schr\"odinger equations, 
we indicated in \cite{yang} that they are all unstable. 
For certain Ginzburg-Landau and coupled-nonlinear-Schr\"odinger type systems, 
Malomed argued that multi-pulse trains exist and are stable by an approximate
method based on the variational principle and effective potential (\cite{malomed1},
\cite{malomed2}). Such results 
need to be viewed with caution due to the approximations involved. The clear
evidence that some multi-pulse waves are stable can be found in the experimental
results on binary fluid convection (\cite{kolodner}) and the numerical results
on subcritical Ginzburg-Landau equations (\cite{brand}). We will investigate
those systems in the near future.

\section*{\hspace{0.1cm} Acknowledgment}
This work was supported in part by the National Science Foundation under the grant
DMS-9622802.

\begin{figure}[p]
\begin{center}
\parbox[t]{8cm}{\postscript{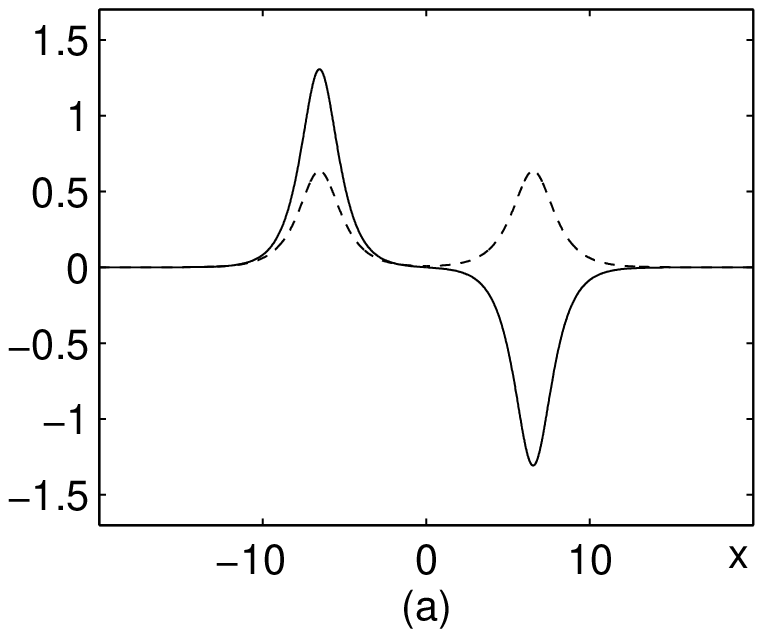}{1.0}}

\vspace{0.5cm}
\parbox[t]{8cm}{\postscript{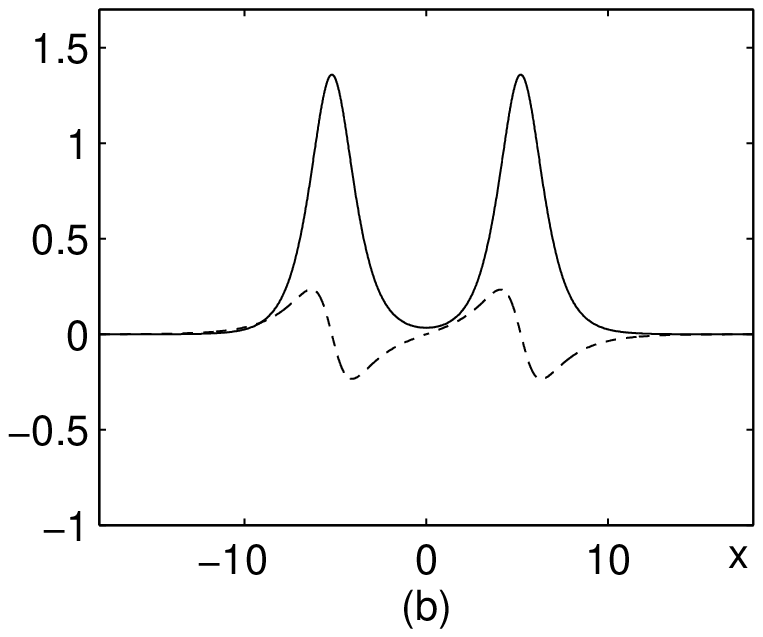}{1.0}}
\vspace{0.5cm}
\parbox[t]{10cm}{
\caption{Solitary waves matched by two wave and daughter wave solutions. 
The solid curves are $r_{1}(x)$, and the dashed curves are $r_{2}(x)$. 
In (a), $(\beta, \omega)=(2/3, 0.85)$ which is close to the curve (\ref{boundary}) with
$m=0$; in (b), $(\beta, \omega)=(2, 0.6)$ which is close to the curve (\ref{boundary})
with $m=1$.}}
\end{center}
\end{figure}

\begin{figure}[p]
\begin{center}
\parbox[t]{8cm}{\postscript{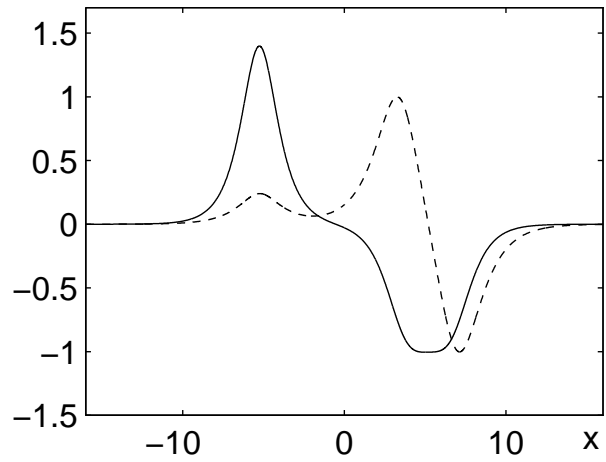}{1.0}}

\parbox[t]{10cm}{
\caption{A solitary wave formed by mixed matching between a wave and daughter wave
solution and another solitary wave of different type. 
Here $(\beta, \omega)=(2/3, 0.78)$, close to curve (\ref{boundary}) with $m=0$.
The solid curve is $r_{1}(x)$, and the dashed curve is $r_{2}(x)$.  } }
\end{center}
\end{figure}

\end{document}